\documentclass{svmult}
\usepackage{graphicx}
\usepackage{subfigure}
\usepackage{amssymb}
\usepackage{amsmath}
\usepackage{stmaryrd}
\usepackage{tikz}

\newcommand{\G}{\mathcal{G}}

\newcommand{\T}{\mathbb{T}}

\newcommand{\SG}{\mathcal{S}_{\G}}
\newcommand{\ST}{\mathcal{S}_{\T}}

\newcommand{\AD}{App}
\newcommand{\DD}{Dis}

\newcommand{\TVG}{\ensuremath{\G=(V,E,\T,\rho,\zeta)}}

\usepackage{makeidx}     
\usepackage{graphicx}    
\usepackage{multicol}    

\makeindex             

\begin{document}

\title*{Selection in Scientific Networks}
\author{Walter Quattrociocchi\inst{1}\and
Frederic Amblard\inst{2}}
\institute{University of Siena
Pian Dei Mantellini 44, 53100, Siena, Italy\\
Tel: +39 0577 233 710\\
\texttt{walter.quattrociocchi@unisi.it}
\and CNRS-IRIT - Universit\'{e} Toulouse 1 Capitole,
2, rue du Doyen Gabriel Marty\\
31042 Toulouse Cedex 9\\
Tel: +33 (0) 561 128 795\\
\texttt{frederic.amblard@univ-tlse1.fr}}
%
%
\maketitle

\begin{abstract}
One of the most interesting scientific challenges nowadays deals with the analysis and the understanding of complex networks' dynamics.
In particular, a major issue is the definition of new frameworks for the visualization and the exploration of the dynamics at play in real dynamic networks.
In this paper we focus in particular on scientific communities by analyzing the ``social part'' of Science through a descriptive approach that aims at identifying the social determinants (e.g. goals and potential interactions among individuals) behind the emergence and the resilience of scientific communities. We consider that scientific communities are at the same time communities of practice (through co-authorship) and that they exist also as representations in the scientists' mind, since references to other scientists' works is not merely an objective link to a relevant work, but it reveals social objects that one manipulates and refers to. In this paper we identify the patterns about the evolution of a scientific field by analyzing a portion of the arXiv repository covering a period of 10 years of publications in physics. As a citation represents a deliberative selection related to the relevance of a work in its scientific domain, our analysis approaches the co-existence between co-authorship and citation behaviors in a community by focusing on the most proficient and cited authors interactions patterns. We focus in turn, on how these patterns are affected by the selection process of citations. 
Such a selection a) produces self-organization because it is played by a group of individuals which act, compete and collaborate in a common environment in order to advance Science and b) determines the success (emergence) of both topics and scientists working on them. The dataset is analyzed a) at a global level, e.g. the network evolution, b) at the meso-level, e.g. communities emergence, and c) at a micro-level, e.g. nodes' aggregation patterns.  
\end{abstract}


\newpage
\section{Introduction}

The evolution of the scientific fields is one of the big issues in Science. On the one hand it deals with the understanding of the factors that play a significant role in such an evolution, not all of them being neither objective nor rational – e.g., the existence of a star system \cite{Wagner2005}, \cite{Newman2001}, \cite{Newman2004}, \cite{Barabasi2002} the blind imitation concerning the citations \cite{MacRoberts96}, the reputation and community affiliation bias \cite{Gilbert77}. On the other hand, having some elements to understand such a dynamics could enable a better detection of the hot topics and of the vivid subfields and how the scientific production is advanced with respect to selection process inside the community itself. 
Among the available data to analyze such a system, a subset of the publications in a given field is the most frequently used such as in \cite{Solla1965}, \cite{Newman2001a}, \cite{Newman2004a}, and \cite{Radicchi2009}. 

The scientific publications correspond to the production of such a system and clearly  identify who are the producers (the authors), which institution they belong to (the affiliation), which funded project they are working on (the acknowledgement) and what are the related publications (the citations), having most of the time a public access to such data explain also a part of its frequent use in the analyses of the scientific field. Classical analyses on these data concern either the co-authorship network (\cite{Barabasi2002, Newman2001}) or the citation network (\cite{Hummon89, Redner05}), more rarely the institutional network (\cite{Powell05}). Moreover, these networks are often considered as static and their structure is rarely analyzed overtime (an exception is the one performed by \cite{Radicchi2009} on Physical Review). In the current paper we introduce two main innovations compared to classical analysis. 
The first one consists in analyzing the scientists' representations of the collaboration structure within the scientific field. Such a representation is captured through the network of cited collaborations, i.e. from a publication we have several references to other papers, each one corresponds to a promotion of the scientists authoring the work. In order to outline the role of this selection process, performed through citations on the scientific advances, we study the evolution of the most cited co-authorship. 
The second innovation deals with the use and analysis of dynamical networks. All the papers are not published at the same time, there is an order that plays a significant role in the structuring and in the advancing of the scientific field. Hence, we decided to take into account such an order while analyzing the cited collaborations. One of the problem when trying to characterize such a structure is that classical indicators from either graph theory or social network analysis cannot be applied directly. Therefore, we used an algebra, the Time-Varying Graphs (TVG) (\cite{CFQS2010}) that enables to take into account the dynamical aspects of networks and allows for the definition of temporal indicators (\cite{ACFQS2010a}) to characterize patterns in evolving structures. 

In the current paper, after presenting the current state of the art concerning the analysis of scientific networks and their results, we present into details the TVG framework as well as the indicators adapted to the dynamical case. Hereafter, we introduce the hep-arxiv dataset we used to make an analysis and we detail the transformation we used in order to obtain the cited collaborations network. In the final part, we present the results of the performed analysis and we conclude our paper with a critical discussion on this method and the next envisioned steps of our work. 

\section{Context}

In this paper we address the problem of characterizing the processes of emergence and self-organization in the scientific systems by selecting a set of indicators able to capture and provide insights about the interaction patterns among scientists (in terms of citations and collaborations). In addition we are interested in outlining how the captured patterns reflect the social factors (goals and related strategies) beyond the scientific production.

In \cite{Newman2001} the network of scientific collaborations, explored upon several databases, shows a clustered and small world structure. Moreover several differences in the collaboration patterns in the different fields studied are captured. Such differences have been deepened in \cite{Newman2004} with respect to the number of papers produced by a given group of authors, the number of collaborations and the topological distances between scientists. Peltomaki and Alava n \cite{peltomaki2006}  propose a new (emulative) model for the growth of scientific network, incorporating bipartition and sub-linear preferential attachment. A model for the self-assembly of creative teams based on three parameters (e.g. team size, the rate of newcomers in the scientific production and the tendency to collaborate with the same group) has been introduced in \cite{Guimera05}. The connectivity patterns in a citation network have been studied in relation to the development of the DNA theory \cite{Hummon89}. Klemm and Eguiluz in \cite{Klemm02} observed that real networks (e.g. movie actors, co-authorship in science, and word synonyms) growing patterns are characterized by a clustering trend that reaches an asymptotic value larger than regular lattices of the same average connectivity. 
In this work we combine both the social processes (i.e. co-authoring on a paper) and their results (i.e. citations) on a temporal perspective. In particular we show how the most proficient authors behave both with respect to co-authorships strategies (the properties of the nodes which they work with) and citations (the productions considered to be relevant by the community).

In the field of social network analysis several works have approached the problem of temporal metrics \cite{Holme05, Kostakos09, KosKW08}. 
Actually, the aim is mainly devoted to capture the intrinsic properties of complex system evolution, that is, capturing and characterizing the dependencies between local behaviors (interactions) and their global effect (emergence) on real networks \cite{Davidsen2002,Mataric92,Woolley1994,amblard01,quattrociocchi2010e}. 
The research approach to social network evolution patterns, at the very beginning was mainly based upon simulations, while in the past few years, due to the large availability of real datasets, either the methodology of analysis and the object of research have changed (\cite{Roth10b,LES07,KosKW08,castellano07,LESK10}). 
In particular, the latter paper states as central problem, for the social networks in general and for the scientific communities networks analysis in particular, the definition of mathematical models able to capture and to reproduce all the properties of dynamical real networks such as the shrinking diameter (\cite{Les05}), or the ``small world'' effect \cite{WATTS99}. Actually instruments and paradigms affording this challenge are mainly based upon stochastic definitions (\cite{Les05b}) or conceptualized as a sequence of static graphs at different times \cite{TSM+09}. 

\section{Tools and Methods}
In this section we first present the empirical dataset explored, then we detail the mathematical framework (TVG) and the related data transformation implemented for the visualization and the analysis of the network evolution.

\subsection{The Empirical Dataset}

The scientific community analyzed in this work has been extracted from the hep-th (High Energy Physics – Theory) portion of the arXiv website, an on-line repository available at http://arxiv.org/. 

The dataset is composed by a collection of papers and therefore their related citations over the period within January 1992 to May 2003. For each paper the set of authors, the dates of the on-line publications on arXiv.org, and the references are provided.
There are 352 807 citations within the total amount of 29 555 papers written by 59 439 authors. 
The broadness of the time window covered allows us to explore the dataset in order to extract, capture and characterize the evolution of the interaction patterns within the community. In particular we will focus on the patterns of the most proficient authors, i.e. the authors that the community, through the selection process of citations, makes emerge.

\subsection{Time Varying Graphs}
The temporal analysis on the dataset is based on Time-Varying Graphs (TVG) formalism, a mathematical framework \cite{CFQS2010} designed to deal with the temporal dimension of networks and to express interactions on interaction-based dynamical systems. 

Consider a set of entities $V$ (or {\em nodes}), a set of relations $E$ between these entities ({\em edges}), and an alphabet $L$ accounting for any property such a relation could have ({\em label}); that is, $E \subseteq V \times V \times L$.  $L$ can contain multi-valued elements. 

The relations (interactions) among entities are assumed to take place over a time dimension $\T$ the {\em lifetime} of the system  which is generally a subset of $\mathbb{N}$ (discrete-time systems) or $\mathbb{R}$ (continuous-time systems). 
The dynamics of the system can subsequently be described by a time-varying graph, or TVG, \TVG, where
\begin{itemize}
\item $\rho: E \times \T \rightarrow \{0,1\}$, called {\em presence function}, indicates whether a given edge or node is available at a given time.
\item $\zeta: E \times \T \rightarrow \mathbb{T}$, called {\em latency function}, indicates the time it takes to cross a given edge if starting at a given date (the latency of an edge could vary in time).
\end{itemize}

Notice that due the nature of the dataset, in the analysis of the current paper the latency function is not considered.

\subsubsection{The underlying graph}
\label{sec:underlying-graph}
\begin{figure}[h]
  \centering
  \begin{tikzpicture}[scale=1.4]
    \tikzstyle{every node}=[draw, circle, minimum size=11pt, inner sep=0pt]
    \path (0,.3) node (a){$a$};
    \path (1, 0) node (b){$b$};
    \path (2.2,0) node (c){$c$};
    \path (3.2,.3) node (d){$d$};
    \tikzstyle{every node}=[font=\scriptsize, inner sep=1pt]
    \draw (a)--node[midway, below, xshift=-2pt, yshift=-2pt]{$[0,1)$}(b);
    \draw (b)--node[midway, below]{$[2,3)$}(c);
    \draw (c)--node[midway, below, xshift=2pt, yshift=-2pt]{$[0,1)$}(d);
  \end{tikzpicture}
  \caption{\label{fig:connected-G} A example of TVG that is not ``connected over time'', although its underlying graph $G$ is connected. {\it Here, the nodes $a$ and $d$ have no mean to reach each other through a chain of interaction.}}
\end{figure}
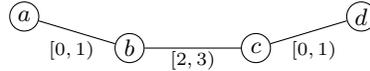

Given a TVG \TVG, the graph $G=(V,E)$ is called {\em underlying} graph of $\G$. This static graph should be seen as a sort of {\em footprint} of $\G$, which flattens the time dimension and indicates only the pairs of nodes that have relations at some time in a given time interval $\T$. It is a central concept that is used recurrently for the analysis in the following sections. In most studies and applications, $G$ is assumed to be connected; in general, this is not necessarily the case.
Note that the connectivity  of $G=(V,E)$ does not imply that $\G$ is connected at a given time instant; in fact, $\G$ could be disconnected at all times. The lack of relationship, with regards to connectivity, between $\G$ and its footprint $G$ is even stronger: the fact that  $G=(V,E)$ is connected does not even imply that $\G$ is ``connected over time'', as illustrated on Figure~\ref{fig:connected-G}.

\subsubsection{Edge-centric evolution}
From an edge point of view, the evolution derives from variations of the availability and the latency over time. TVG defines the {\em available dates} of an edge $e$, noted $\I(e)$, as the union of all dates at which the edge is available, that is, $\I(e)= \{t \in \T : \rho(e,t)=1\}$. 
Given a multi-interval of availability $\I(e)=\{[t_1,t_2)\cup[t_3,t_4)...\}$, the sequence of dates $t_1,t_3,...$ is called {\em appearance dates} of $e$, noted $\AD(e)$, and the sequence of dates $t_2, t_4,...$ is called {\em disappearance dates} of $e$, noted $\DD(e)$. Finally, the sequence $t_1, t_2, t_3,...$ is called {\em characteristic dates} of $e$, noted $\ST(e)$.

\subsubsection{Graph-centric evolution}

From a global standpoint, the evolution of the system can be given by a sequence of (static) graphs $\SG=G_{1}, G_{2}..$ where every $G_i$ corresponds to a static {\em snapshot} of $\G$ such that $e\in E_{G_i} \iff \rho_{[t_i,t_i+1)}(e)=1$, with two possible meanings for the $t_i$s: either the sequence of $t_i$s is a discretization of time (for example $t_i=i$); or it corresponds to the set of particular dates when topological events occur in the graph, in which case this sequence is equal to $sort(\cup\{\ST(e): e \in E\})$. In the later case, the sequence is called  {\em characteristic dates} of $\G$, and noted $\ST(\G)$. 

\paragraph{Subgraphs of a time-varying graph}
Upon this framework it is possible to define a temporal subgraph $\G'$ by restricting the lifetime $\T$ of $\G$, and leading to the graph $\G'=(V,E',\T',\rho',\zeta')$ such that 

\begin{itemize}
\item $\T' \subseteq \T$
\item $E' = \{e \in E : \exists t \in \T' : \rho(e,t)=1 \wedge t+\zeta(e,t)\in \T'\}$
\item $\rho'(e,t)=\rho(e,t)$ for any $e\in E'$ and $t\in \T'$
\item $\zeta'(e,t)=\zeta(e,t)$ for any $e\in E'$ and $t\in \T'$
\end{itemize}

\subsection{Expliciting Interactions}
As social interaction in scientific communities depends principally upon competitions and collaborations among authors and groups, in the analysis we want to capture both the resulting emerging effects caused by these two opposite motivations and how they are expressed in terms of connectivity and citations patterns. 

The dataset analyzed in the current paper presents two explicit interactions: the papers' co-authorships and the citations between papers.  

The former can be influenced by authors' proximity (working in the same institution or in the same scientific field), by the nature of the problems addressed, and often by the complementarity between scientists' skills (in order to cover all the aspects addressed in a scientific work).  
The latter, in turns, is affected by the authors' background knowledge and by the scientific histories of the addressed topics (i.e. milestones, fundamental contributions, etc.). 
In addition, there is an implicit level of interaction that depends upon the goals behind each research paper: the quality and, at the same time, the necessity to be highly cited (competition). Hence, often both the collaborations' and citations' strategies are optimized in order to have the highest impact with respect to the problem addressed and to collect the highest number of citations.

\subsubsection{The Interaction Network}
In order to exploit and express all the descriptive potential of the dataset's social interaction domain, we approach the data transformation in order to explicit the co-authorships and cited collaboration patterns. 
We represent the dataset as an undirected graph, namely {\em interaction network}, having as nodes the authors, weighted links representing the co-authorship on a paper, and when a paper is cited by another work, the links' weights, connecting the authors of the referenced paper, are incremented.

More formally, the graph of the {\em cited co-authorships} is defined as   \TVG  on a discrete time and with the value of the latency function$\zeta$ fixed to 0. 
Here the elements $v \in V$ are the authors, the set of edges $E \subseteq V \times V \times L$ represents the collaborations $L$ on a paper's production. The nodes appear on the graph the first time a paper they wrote has been published, and the interaction $L$ is weighted with a variable $w_i$, namely the {\em strength value} of a collaboration, that is incremented of one for each citation received by a given couple of nodes $(u,v)$. 

In the paper we analyze and report on the behaviors and on the interaction's strategies within the most cited authors' network, such a graph, namely $G_i$, is a subset of the global interaction network \TVG. 
In particular the nodes considered in the analysis are only the authors having links' strength values higher than 150, that is, all the groups having more than 150 citations on a work. Such a network in its maximal expansion, during the 10 years temporal window observed, is composed by 12 583 nodes and 84 512 edges.

\section{Results}
The results drawn from the analysis are presented and discussed in this section. The presentation is structured in order to provide to the reader a threefold top-down perspective on the emergent processes characterizing the evolution of the scientific network. 
First, we provide an outline of the global network dynamics, then we show the meso and micro levels of the interactions network by presenting the community formation patterns and the evolution of nodes' interconnections (cited co-authorships). For each metric used in the analysis the related definition in terms of the time-varying graph formalism is provided.

\subsection{The Network}
From a global point of view, the evolution of the interaction network is characterized by computing a collection of temporal indicators, defined in the TVG formalism, at different time intervals - e.g., the evolution of the clustering coefficient,  the temporal trend of the average degree, of the average path length, of the degree power law, of the modularity and of the density \cite{ACFQS2010a}. 

These values are computed on the {\em temporal subgraph sequence}  $S_n$  of the interaction network formed by the most cited groups (authors with more than 150 citations). Each element $s_i$ of the sequence is a time-varying graph defined as $G_{\T^{i}} = (V(\T^{i}),E(\T^{i}))$ with $i$ being a given time interval such that 

\begin{itemize}
 \item $E(\T^{i}) :\{e \in E | \rho(e,t) = 1 \forall t\in \T^{i} \}$  
 \item $V(\T^{i}): \{v \in V | \exists y \in V \wedge (x,y) \in E(\T^{i})\}$
\end{itemize}

Note that the indicators for each element of the sequence are computed over the underlying graph (see section \ref{sec:underlying-graph}) at a time interval of one year. 

\subsubsection{The Phase Transition}

The most important element that emerges at this level of observation of the network evolution is a phase transition occurring within 1999 and 2000.

\begin{figure}[h!]
 \centering
 \includegraphics[width=70mm]{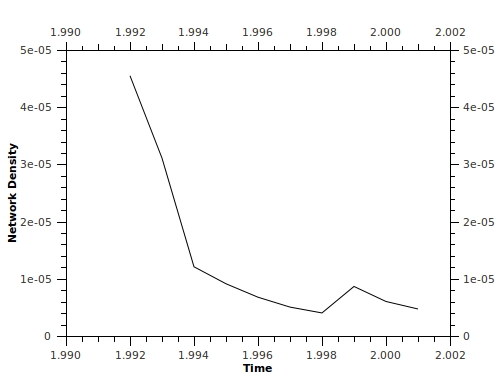}
 \caption{Density}
 \label{fig:Density}
\end{figure}

\paragraph{Density Evolution}
A dense network is one in which the number of edges is close to the maximal number of edges.
Figure \ref{fig:Density} shows the density values for each element of the temporal subgraphs sequence of the interaction network.

The density trend starts with very low values and then an increase of the graph sparsity during the evolution occurs with a very low counter-trend during the period of 1999 and 2000.

\paragraph{Modularity Evolution}

The modularity, introduced by \cite{blondel08}, measures how a network can be decomposed into subparts, i.e. classically finding partitions within a graph. It allows to look at communities at different resolutions in order to detect the network structure and its evolution. Given two nodes $u$ and $v$ and the number of edges between them, in order to compute the modularity, we have to find the right partitioning of the network in a number of groups (some recent algorithms enable however to overcome the partitioning constraint and enable to detect overlapping communities \cite{Palla2007} \cite{Cazabet2010}). Note that such partition is temporal, i.e. related to a temporal subgraph of the interaction network.

\begin{figure}[h!]
 \centering
 \includegraphics[width=70mm]{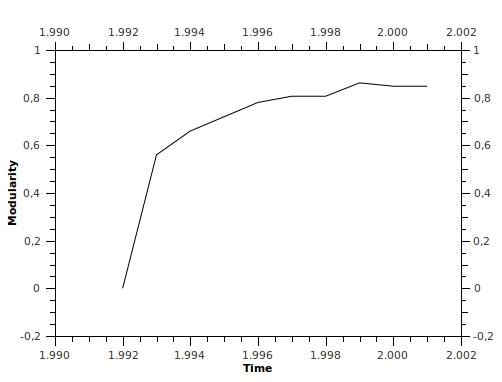}
 \caption{Modularity}
 \label{fig:Modularity}
\end{figure}

The modularity for each subgraph of the interaction network's subgraphs sequence is shown in Figure \ref{fig:Modularity}.

It shows how the quality of a division of a network into modules or communities evolves during time. The trend of these values says that there is an increase of dense internal connections between the nodes within modules but only sparse connections between different submodules. 
Hence the communities tend to remains separated, only few nodes act as bridges between different groups. The growing rate of the modularity is characterized by an  increase until 1993, then it reaches its highest values during the 1999-2000 interval, but through a smoothed increasing rate. 
As far as we can see by the modularity evolution, the interconnections among separated groups of authors starts in 1993, then their interconnection continues, but with a gradual rate. 

\paragraph{Average Clustering Coefficient Evolution}
In order to capture the global nodes' interconnections patterns we show the clustering coefficient evolution during the time window observed.
The {\em clustering coefficient} $C(v_i)$ is the proportion of edges between the node $v_i$ within its neighborhood divided by the number of edges that could potentially exist between them  \cite{WATTS99}.

\begin{figure}[h!]
 \centering
 \includegraphics[width=70mm]{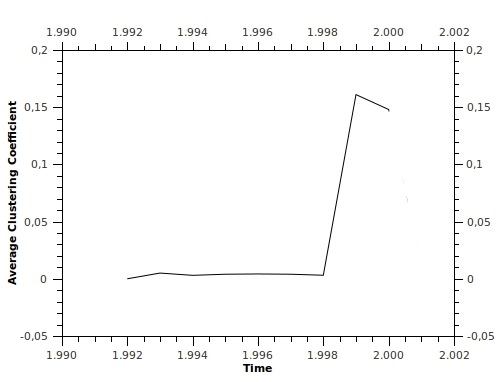}
 \caption{Clustering}
 \label{fig:Clustering}
\end{figure}

Figure \ref{fig:Clustering} a phase transition during this period that was not present neither in the density nor in the modularity evolution. The chart suggests that there is a trend among authors to remain clustered in tightly knit groups. Such a tendency increases between 1999 and 2000.

\subsection{A Matter of Interconnections}

\begin{figure}[h]
 \centering
 \includegraphics[width=70mm]{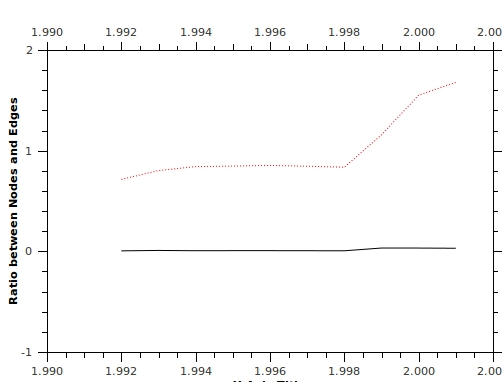}
 \caption{Average connectivity: average number of edges per node for the entire network in the dataset (in black) and for the network of the most proficient authors (in red)}
 \label{fig:Ratio}
\end{figure}

As shown in  Figure \ref{fig:Ratio}, where the trend of the ratio between nodes and edges of both the whole interaction network and the network of the most proficient scientists are depicted, the phase transition process, evinced in the clustering coefficient evolution in Figure \ref{fig:Clustering} , is not caused by an increase of the number of authors in the period between 1999 and 2000, neither it is a pattern common to the whole dataset. We can see that in the collaborations network of all authors (in black) there are no particular change in 1999, while the collaboration network of the most proficient authors shows a phase transition, caused by the increasing number of connections (i.e. collaborations) among the most cited nodes.

In Table \ref{tab:otherindicators} we display the evolution of the average degree, the average path length and of the power law degree within the temporal window observed. 
As for the previous indicators these values are computed on the underlying graph of the interaction network of the most proficient scientists.

\begin{table}
\begin{center}
 \begin{tabular}{|c|c|c|c|}
\hline 
Year & Average Degree & Average Path Len & Power Law\tabularnewline
\hline
\hline 
1992 & 0,0095 & 1 & 0\tabularnewline
\hline 
1993 & 0,0176 & 1 & -1,386\tabularnewline
\hline 
1994 & 0,012 & 1 & -1,79\tabularnewline
\hline 
1995 & 0,0135 & 1,16 & -2,16\tabularnewline
\hline 
1996 & 0,132 & 1,13 & -2,27\tabularnewline
\hline 
1997 & 0,0118 & 1,12 & -2,5\tabularnewline
\hline 
1998 & 0,106 & 1,12 & -2,5\tabularnewline
\hline 
\textbf{1999} & \textbf{0,066} & \textbf{3,92} & \textbf{-5,08}\tabularnewline
\hline 
\textbf{2000} & \textbf{0,64} & \textbf{3,79} & \textbf{-5,27}\tabularnewline
\hline 
2001 & 0,6 & 3,82 & -5,25\tabularnewline
\hline
\end{tabular}
\label{tab:otherindicators}
\caption{Other interaction network's measurements}
\end{center}
\end{table}
In bold the values when the phase transition occurs. Neither the average path length, indicating the average distances among nodes, the power law degree, measuring how closely the degree distribution of a network follows a power-law scale and the evolution of average degree, counting the average number of connections at each node, are immune to the phase transition.

\subsection{Communities Emergence}

As the phenomena behind the change phase transition are mainly caused by the evolution of the interconnections among the nodes of the cited co-authorship graph. In this section we outline the connectivity patterns at a community level.

\subsubsection{Beyond Preferential Attachment}

\begin{figure}[h!]
 \centering
 \subfigure[Several separated connected components]
   {\includegraphics[width=45mm]{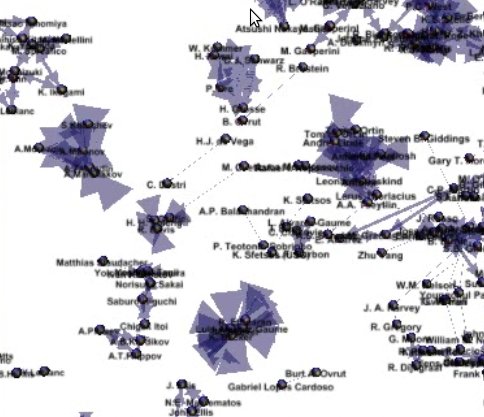} \label{fig:6}}
 \hspace{1mm}
 \subfigure[that start to connect with each other]
 {\includegraphics[width=51mm]{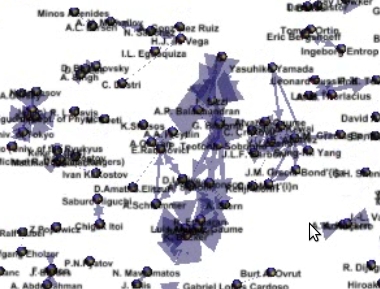} \label{fig:7}}
 \caption{Connections within the islands}
\label{fig:comm1}
\end{figure}

We start with the introducing of a sequence of screen-shots showing the nodes' aggregation patterns. The pictures, obtained with the gephi platform (\cite{Gephi09}), refer to the biggest community within the most proficient authors' network during the phase transition period (i.e. 1999-2000). 
The sequence of snapshots is the interaction patterns behind the formulation of the ``String Theory'' and of its consequent developments. 
In fact, among the nodes in this portion of the network there are E.Witten, N.Seiberg and so forth playing the role of attractors. Authors in the same community start to join their group, it is a picture of what is beyond the preferential attachment - e.g.,mechanism used to explain the power law degree distributions in social networks - when the system is goal-driven.
At the beginning (Figure \ref{fig:6}) there are several separated components, that start to mix by co-authoring papers (Figure \ref{fig:7}).

\begin{figure}[h!]
 \centering
 \subfigure[The number of connections within authors continues to increase.]
   {\includegraphics[width=51mm]{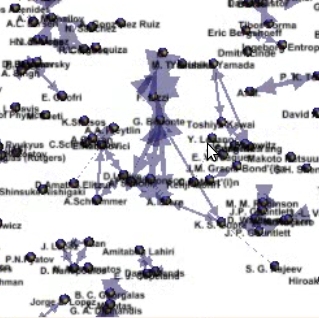} \label{fig:8}}
 \hspace{1mm}
 \subfigure[The maximum level of connectivity is reached.]
 {\includegraphics[width=45mm]{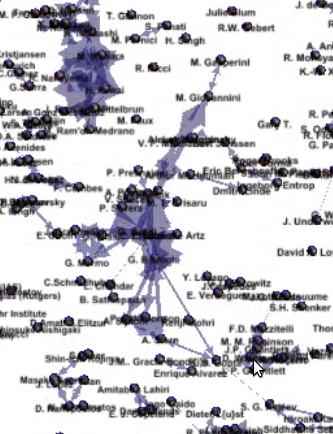} \label{fig:9}}
 \caption{The growing phase of interaction among authors}
\label{fig:comm2}
\end{figure}

This group starts to play the role of attractor with respect to the neighboring nodes as it is shown in Figure \ref{fig:8} until the maximum level of connections in the group is reached in Figure \ref{fig:9}.

Note that in Figure \ref{fig:comm1} and in Figure \ref{fig:comm2} the links are emphasized in proportion to the number of citations received by the papers' authors. The component in the center is highly cited and is playing the role of attractor on the neighboring nodes (authors). It is a goal-driven preferential attachment due to the number of citations (representing the emergence through selection) to a given group, that in terms of the goal of any scientific community clearly evinces a strategy oriented to the community belonging and to the couplage between topics and sub-communities, authors tend to join highly cited groups to satisfy both the quality and the possibility to be highly cited requirements.
Moreover, considering that at the beginning there are several separated groups, the phenomenon can be interpreted as a three-fold process with a first phase as the exploration of ideas by means of separated works afforded by separated groups, a second one when a part of the ideas explored starts to be cited more than the others, and a third one when authors tend to join groups that have produced highly cited works. 

Such a process presents the phases of the the natural selection, e.g. the exploration, the selection and migration. But here a) such a (social) selection produces self-organization because it is played by a group of individuals which act, compete and collaborate in order to advance science and b) it determines the success (emergence) of a topic and of the scientists working on it.

\subsubsection{Community Evolution}

In this section, we are going to quantify in a temporal fashion the patterns observed in the previous section. Table \ref{table:metrics} summarizes the network evolution by means of a) basic indicators, e.g. the number of nodes, the number of edges and the community's diameter) and b) aggregated indicators, e.g. the cyclomatic number, the alpha, beta, and gamma index. 
The {\em cyclomatic number} counts the number of cycles on the graph, its magnitude characterizes the development of the nodes' accessibility. The {\em alpha index} is the ratio between the number of cycles in the graph and their possible maximum value. The range of the alpha index spread within 0 to 1, that are from no cycles to a completely interconnected network.
The {\em beta index}, is a simple measure of connectivity. It relates to the total number of edges to the total number of nodes. The higher the value, the greater the connectivity. The {\em gamma index} measures the ratio between the number of edges on the network and the maximum number of possible edges among nodes. The gamma index spreads within 0 and 100, respectively indicating the minimum and the maximum number of edges between nodes.

As we can see from the evolution of these parameters, the aggregation pattern among separated components is evident for each one of the metric proposed. In terms of nodes that join the community and their mutual connections, the diameter over time passes through a phase of expansion and then tends to stabilize.

\begin{center}
\begin{table}[h!]
 \begin{tabular}{|c|c|c|c|c|c|c|c|}
\hline 
Measures & April 00 & October 00 & April 01 & October 01 & April 02 & October 02 & April 03\tabularnewline
\hline
\hline 
Vertices:             & 23 & 51 & 65 & 66 & 67 & 70 & 72\tabularnewline
\hline 
Edges:                & 29 & 75 & 99 & 100 & 106 & 110 & 114\tabularnewline
\hline 
Diameter:             & 6 & 10 & 10 & 10 & 8 & 8 & 8\tabularnewline
\hline 
Cyclomatic:  & 17 & 25 & 35 & 35 & 40 & 41 & 43\tabularnewline
\hline 
Alpha:  & 0,73 & 0,02 & 0,017 & 0,016 & 0,018 & 0,017 & 0,017\tabularnewline
\hline 
Beta: & 1,69 & 1,47 & 1,52 & 1,51 & 1,58 & 1,57 & 1,58\tabularnewline
\hline 
Gamma: & 61,9 & 51,02 & 52,38 & 52,08 & 54,3 & 53,92 & 54,28\tabularnewline
\hline 
\end{tabular}
\label{table:metrics}
\caption{network measurement of the biggest community}
\end{table}               
\end{center}

\section{Conclusions}
In this paper we analyse the behavior of the most cited authors in a collection of papers extracted from the on-line repository of arXiv. We were captured and characterized the evolution of the network in terms of interactions (citations and co-authorships) within a given scientific community. 

The temporal dimension and the metrics used for the analysis were formalized using Time-Varying Graphs (TVG), a mathematical framework designed to represent the interactions and their evolution in dynamically changing environments.

The analyses, focusing on the cited co-authorship's patterns, have been performed at different levels. At a global level with respect to the network evolution; at a meso-level with respect to the communities aggregation patterns and finally at a micro-level, characterizing the accessibility trend of the biggest community in the network. 
Each level has shown a particular trend, that, as far as we can see on the analysis, is given by a phase transition within 1999-2000. Such trend is caused by an increase of the interconnections among nodes in the network. It is a sort of preferential attachment driven by the number of citations received by a given group, that in terms of the goal of any scientific community clearly evinces a strategy oriented to the community belonging, authors tends to join highly cited groups.
This fact together with the fact that at the beginning there are several separated groups can be interpreted as a three-fold process: the first phase is the exploration of ideas by means of works, once some ideas start to be cited more than others, then, finally authors tend to join groups that have produced highly cited works. 
Such a process is similar to the natural selection, in fact it passes through  the exploration, the selection and migration phase. But here the selection is performed by individuals in a goal oriented  environment and such a (social) selection produces self-organization because it is played by a group of individuals which act, compete and collaborate in order to advance Science. In addition, the social selection determines the emergence of a topic and of the scientists working on it by determining the preferential attachment patterns.
In the next future we are going to outline the behavior of the most proficient scientist in terms of their aggregation patterns, and on how their works are diffused within the community, that is, characterizing the reasons behind the selection process beyond the network evolution.
Such aspects will be addressed both with new analyses on different datasets and by means multi-agent simulations. The former stream will be devoted to the definition of new patterns, the latter will be used for the understanding of how changing some parameters of the network influences the evolution, and consequently the quality, of the scientific production.

\section{Acknowledgement}
This work was partially supported by the Future and Emerging Technologies programme FP7-COSI-ICT of the European Commission through project QLectives (grant no.: 231200).

\bibliographystyle{plain}
\bibliography{dtn}

\printindex
\end{document}